\documentclass[conference]{IEEEtran}
\usepackage[flushleft]{threeparttable}
\usepackage{listings}
\usepackage{subcaption}
\usepackage{url}
\usepackage[hidelinks]{hyperref}

\IEEEoverridecommandlockouts
\usepackage{cite}
\usepackage{amsmath,amssymb,amsfonts}
\usepackage{algorithmic}
\usepackage{graphicx}
\usepackage{textcomp}
\usepackage{xcolor}
\def\BibTeX{{\rm B\kern-.05em{\sc i\kern-.025em b}\kern-.08em
    T\kern-.1667em\lower.7ex\hbox{E}\kern-.125emX}}

\newcommand{\etal}{\emph{et al.}}

\begin{document}

\title{Towards Better Quality of Experience in\\HTTP Adaptive Streaming}


\author{Babak Taraghi\\
\IEEEauthorblockA{\small{\textit{Christian Doppler Laboratory ATHENA}}\\
\small{\textit{Institute of Information Technology (ITEC)}}\\
\small{\textit{Alpen-Adria-Universität Klagenfurt}}\\
Klagenfurt, Austria\\
babak.taraghi@aau.at}
\and
\IEEEauthorblockN{Selina Zoë Haack}
\IEEEauthorblockA{\small{\textit{Christian Doppler Laboratory ATHENA}}\\
\small{\textit{Institute of Information Technology (ITEC)}}\\
\small{\textit{Alpen-Adria-Universität Klagenfurt}}\\
Klagenfurt, Austria\\
zohaack@edu.aau.at}
\and
\IEEEauthorblockN{Christian Timmerer}
\IEEEauthorblockA{\small{\textit{Christian Doppler Laboratory ATHENA}}\\
\small{\textit{Institute of Information Technology (ITEC)}}\\
\small{\textit{Alpen-Adria-Universität Klagenfurt}}\\
Klagenfurt, Austria\\
christian.timmerer@aau.at}
}

\maketitle

\begin{abstract}
HTTP Adaptive Streaming (HAS) is nowadays a popular solution for multimedia delivery. The novelty of HAS lies in the possibility of continuously adapting the streaming session to current network conditions, facilitated by Adaptive Bitrate (ABR) algorithms. Various popular streaming and Video on Demand services such as Netflix, Amazon Prime Video, and Twitch use this method. Given this broad consumer base, ABR algorithms continuously improve to increase user satisfaction. The insights for these improvements are, among others, gathered within the research area of Quality of Experience (QoE). Within this field, various researchers have dedicated their works to identifying potential impairments and testing their impact on viewers' QoE. Two frequently discussed visual impairments influencing QoE are stalling events and quality switches. So far, it is commonly assumed that those stalling events have the worst impact on QoE. This paper challenged this belief and reviewed this assumption by comparing stalling events with multiple quality and high amplitude quality switches. Two subjective studies were conducted. During the first subjective study, participants received a monetary incentive, while the second subjective study was carried out with volunteers. The statistical analysis demonstrated that stalling events do not result in the worst degradation of QoE. These findings suggest that a reevaluation of the effect of stalling events in QoE research is needed. Therefore, these findings may be used for further research and to improve current adaptation strategies in ABR algorithms.
\end{abstract}


\section{Introduction}
\label{section:intro}
HAS has been disrupting the industry since its first release by Move Networks. Various researchers are focusing on optimizing the adaptation logic that brought HAS its popularity. A significant part of the development of HAS is credited to Quality of Experience (QoE) research, which aims to improve the consumers’ experience~\cite{ni2011flicker, muellerlederer, rodriguez2014impact, babak, taraghi2021understanding}. These studies aim to identify parameters that positively and negatively impact the users’ experience. A stalling event happens when the player buffer is empty and the player can not continue the playback, indicated often with a spinning wheel. The quality switch is the player function in which the media representation will be changed to continue the playback with lower or higher media quality. Following the focus of this paper, three main findings should be pointed out:

\begin{itemize}
    \item Stalling events are generally assumed to have the worst impact on QoE~\cite{seufertcasas, bokani2015optimizing}
    \item Multiple quality switches have a negative effect on QoE~\cite{ni2011flicker, zink2005layer}
    \item High amplitude quality switches decrease the QoE~\cite{ni2011flicker, rodriguez2014impact}
\end{itemize}

Nevertheless, no research has compared multiple quality switches, high amplitude quality switches, and stalling events, which have been proven to impact QoE negatively. This paper has conducted evaluations and analyses to provide further insights into these areas.

Following the introduction, the paper is divided into the following sections: Section~\ref{section:related} provides an overview of the background of this study. Section~\ref{section:method} focuses on the research methodology and study setup. And in Section~\ref{section:results} we present the paper's analysis of results and findings. Finally, in Section~\ref{section:conclusions} we conclude the paper with possible future works.

\section{Background Overview}
\label{section:related}

\subsection{HTTP Adaptive Streaming}
HAS uses the HTTP protocol to deliver data via the network. The usage of this protocol comes with the significant advantage of accessing the whole architecture built around it. That includes proxies, which redirect the internet traffic, caches, which improve performance, and Content Delivery Networks (CDNs), which store content closer to the end-user~\cite{muellerlederer}.\looseness=-2

The Moving Picture Expert Group developed Dynamic Adaptive Streaming over HTTP (MPEG-DASH) over two years. In 2011, after submitting multiple drafts, it was eventually published as an international standard (ISO/IEC 23009-1), and since its publication, it has been revised numerous times.

\subsection{Quality of Experience}
\begin{figure*}[tp]
    \includegraphics[width=\linewidth]{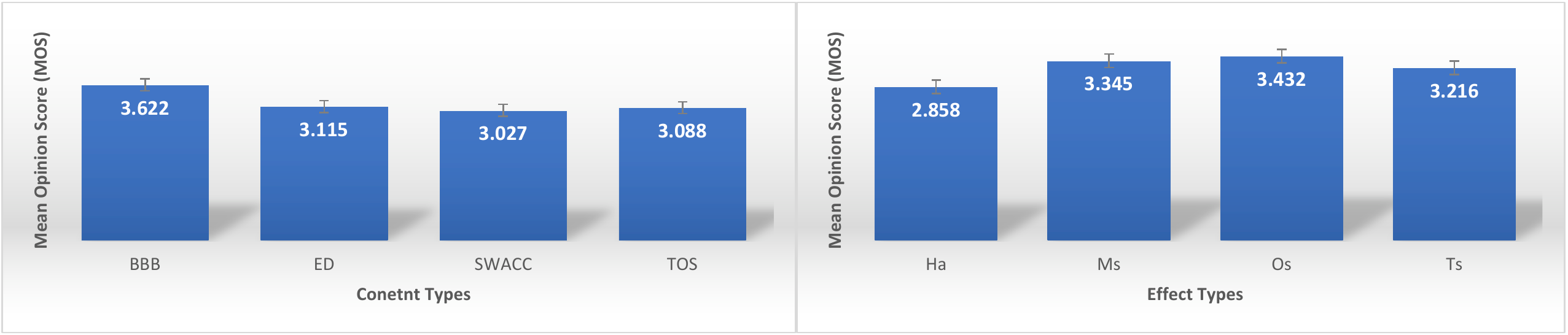}
    \caption{Subjective Study Results with Standard Error Bars between the Content Types (Left) and between the Effect Types (Right)}
    \label{fig:effectAndContent}
\end{figure*}

Before the term QoE was introduced, various kinds of research had been conducted under the roof of Quality of Service (QoS). A good QoS score, however, does not necessarily translate to user satisfaction. Therefore a shift in research from a quantifiable network to a subjective user-centered approach was necessary~\cite{Mller2014}. European Network on Quality of Experience in Multimedia Systems and Services (Qualinet) defines QoE as follows:

"Quality of Experience (QoE) is the degree of delight or annoyance of the user of an application or service. It results from the fulfillment of his or her expectations with respect to the utility and/or enjoyment of the application or service in the light of the user’s personality and current state."~\cite{qualinet}

\section{Methodology and Study Setup}
\label{section:method}
\subsection{Crowdsourced Approach}
The decision to conduct this study via crowdsourcing was mainly based on two reasons: it generally requires fewer resources and the ongoing COVID-19 pandemic. Selecting, inviting, and training participants in a lab environment requires many resources such as time, lab space, and equipment. In crowdsourcing, participants can be quickly recruited from a pool of workers. In addition, they can conduct the study at their convenience with their equipment, which makes it less resource extensive for participants and researchers. Furthermore, at the time of conducting this research, the COVID-19 pandemic was still ongoing. Therefore, inviting people to a closed lab environment would have been socially irresponsible. As the usage of crowdsourcing platforms in QoE studies is still limited, this approach will be explained shortly.
Crowdsourcing is generally used to outsource tasks to a global pool of workers with different skills and backgrounds. There are various advantages such as a larger diversity in test subjects, quick and efficient turnover, and, as mentioned earlier, less resource extensive than lab testing. That includes time, monetary and technological resources. There are different platform types as well as task types in crowdsourcing. Independent of the platform type, crowdsourcing platforms mainly function as collectors and distributors of tasks. Tasks such as the sequence rating used in this study are categorized as so-called microtasks that only take a couple of minutes to conclude. To ensure that a suitable user base is attracted to the tasks, most platforms can further filter for different parameters such as system devices or specific skills.
While crowdsourcing offers many advantages, it also introduces various challenges to the researcher. A common challenge is the task's potential misinterpretation due to unclear instruction and the lack of a test moderator. Therefore, the evaluation instruction should be as clear and detailed as possible.
Another issue that should be considered in crowdsourcing is the effect of the incentive. Some participants might cheat or not fulfill the task correctly to gain the incentive faster. To prevent and filter the submission of such results, Hossfeld~\etal proposes different techniques in their paper about crowdsourcing best practices \cite{qoebestpractice}. One is to implement "gold standard data". This refers to introducing questions to which the answer is already known. This method allows verifying whether the participants paid attention to the initial instructions and are answering the study questions accordingly and truthfully. There have been different works about the implementation of gold standard data. Some suggest instantly revealing the correctness of the data \cite{le2010ensuring} while others use it in the post-evaluation procedure \cite{ipeirotis2010quality, snow2008cheap}. In addition, proper task design was also shown to impact cheat prevention. A study revealed that participants were less likely to cheat if the task was designed so that cheating would require the same amount of time as conducting the study properly \cite{kittur2008crowdsourcing}.
Best practices also suggest considering technical implementations, such as preventing unwanted impairments. It is advised to implement a measure that could cache the sequences before playback to ensure that participants with lower bandwidth accessibility would not experience any unplanned impairments \cite{qoebestpractice}. 
In addition, it is recommended to consider the selection of web server size as crowdsourcing platforms attract thousands, sometimes millions, of workers; therefore, it could be possible that hundreds of people access the study simultaneously. Ultimately, regarding implementation, it is advised to use a standard scripting language such as JavaScript to ensure system and device compatibility as some browsers might not support uncommon ones \cite{qoebestpractice}.

In this study, the platform Amazon Mechanical Turk (MTurk) was used. The specific functionalities of MTurk are further explained in Section~\ref{mturk}.

\subsection{Dataset Selection}
\label{dataselect}
The sequences for the study were chosen from four different original sources (Table~\ref{tab:sequences}). The first one is Big Buck Bunny, an animated movie from the Blender Foundation which stars animals as main characters and displays a moderate amount of motion. Furthermore, the short film includes large homogeneous backgrounds and bright settings. Secondly, Elephants Dream, also part of the Blender portfolio, portrays a man and a woman walking through a strange world, potentially displaying the growing insanity of the world. The movie is slow-paced and takes mainly place in a dark setting. The third test sequence was selected from The Swiss Account, a self-made documentary about bouldering that includes many close-ups and detailed shots. Last but not least, a sequence was taken from Tears of Steel, a high-motion action movie with various special effects. The movie starred real actors alongside animated machines.
The variety in content was chosen to reduce the potential impact of content type, which can influence the ratings if a participant has a general like or dislike for one specific content type. Displaying different sequences also decreases the reduction in interest, resulting from watching the same content multiple times. Each of the prepared sequences was 30 seconds in length and therefore only an excerpt of the original content. The exact timestamps and the order in which the sequences were displayed during the subjective study are summarized in Table~\ref{tab:sequences}. All of the selected videos are accessible under the Creative Commons license.

\begin{table}[tp]
\centering
\caption{Test Sequences}
\label{tab:sequences}
\begin{tabular}{|c|c|c|c|}
\hline
    \textbf{Name} & \textbf{Source} & \textbf{Timestamp} & \textbf{Order}\\
\hline
    Big Buck Bunny~\tnote{1} & Blender Foundation & 01:55 - 02:25  & 1 \\ 
    Big Buck Bunny & Blender Foundation & 05:45 - 06:15  & 2 \\ 
    Elephants Dream~\tnote{2} & Blender Foundation & 04:00 - 04:30  & 3   \\ 
    The Swiss Account~\tnote{3} & Louder Than 11 & 41:40 - 42:10 & 4  \\ 
    Tears of Steel~\tnote{4} & Blender Foundation & 07:25 - 07:55 & 5 \\ 
\hline
\end{tabular}
\begin{tablenotes}
    \item[1] Big Buck Bunny, retrieved from \cite{bbb}
    \item[2] Elephants Dream, retrieved from \cite{ed}
    \item[3] The Swiss Account, retrieved from \cite{swacc}
    \item[4] Tears of Steel, retrieved from \cite{tos}
\end{tablenotes}
\end{table}

\subsection{Sequence Preparation}
\label{seqprep}
Following the best practice framework of Hossfeld~\etal the sequences were prepared in advance \cite{qoebestpractice}. After selecting the original content introduced in the previous section, the source content was checked via \textit{ffprobe}. Afterwards, they were encoded via \textit{FFmpeg} with the following command:

\begin{lstlisting}[basicstyle=\small, breaklines=true, language=Bash]
$ ffmpeg -i input_file.mp4 \
-map 0:v -map 0:v -map 0:v -map 0:a:0 \
-c:a aac -c:v libx264 -x264opts 'keyint=24:min-keyint=24:no-scenecut' \
-dash_segment_type mp4 -seg_duration 5 -use_timeline 0 
-use_template 1 -keyint_min 24 -g 24 \
-init_seg_name \$Bandwidth\$/init.m4s 
-media_seg_name \$Bandwidth\$/seg-\$Number%01d\$.m4s \
-b:v:0 4300k -s:v:0 1920x1080 \
-b:v:1 1050k -s:v:1 640x480 \
-b:v:2 235k -s:v:2 320x240 \
-b:a:3 128k -ar 48000 -ac 2 \
-adaptation_sets "id=0,streams=v id=1,streams=a" -f dash -r 24 manifest.mpd
\end{lstlisting}

Initially, the path to the original file is input, and four streams are mapped. The first three represent video streams to enable bitrate switching, and the last one represents the audio stream. The audio stream is encoded with Advanced Audio Coding (AAC), which was introduced shortly in Sub-section~\ref{codecs} of the Terminology Section (Section~\ref{terminology}). \textit{Libx264}, a platform-independent encoder for the H.264/ MPEG-4 AVC video format, was used for the video streams. In addition, settings were customized to set the maximum size of a Group of Pictures (GOP) to 24 and to prevent scene cuts. The prevention of scene cuts ensures that the encoder does not use its automated I-frame detection algorithm to add additional I-frames to the encoding. As I-frames are more expensive, it is recommended to limit the amount of inserted I-frames when encoding streams \cite{furht2003handbook}.
The "dash-segment-type mp4" sets the file format to ISOBMFF, the ISO/IEC standard format for media segments \cite{weorgisobmff, isombff}. In addition, the segment duration is set to 5, which means the algorithm could decide to switch to a different quality every 5 seconds. Furthermore, the segment template was enabled while the segment timeline was disabled. These settings ensure that the previously set segment duration is treated as the actual duration time and not as the minimum duration time \cite{ffmpegdocumentation}. In addition, two templated variables in \textit{dash} are set. Line 6 specifies the naming convention for the initialization segment, whereas line 7 describes the naming convention for the following media segments. Media segments are set to be named in descending order.
The following four lines define the bitrate levels for the defined streams. Stream 0, 1, and 2 represent the video streams and are mapped with the corresponding bitrate and frame size. The videos were encoded at three different resolutions: 1080p, 720p and 240p. The bitrate ladder used in this study was based on a standard by Netflix \cite{netflixbitrate}. Stream 3 corresponds to the audio and describes the bitrate and two additional parameters: the audio sample rate in hertz and the number of audio channels. The exact specifications of each stream, such as bitrate, resolution, codec, and segment length, can be seen in Table~\ref{tab:bitrate}.
The last two lines focus on DASH-specific settings. First, the mapped streams are divided into two adaptation sets: audio and video. Following, the DASH muxer is called upon to create the individual segments and the manifest file needed for \textit{dash} playback \cite{ffmpegdocumentation}.

\begin{table}[tp]
\centering
\caption{Bitrate Encoding}
\label{tab:bitrate}
\begin{tabular}{|c|c|c|c|c|c|}
\hline
Stream & Bitrate & Resolution & Type & Codec & Length\\
\hline
Stream0 & 4300kbps & 1920x1080 & video & x264 & 5s\\
Stream1 & 1050kbps & 854x480 & video & x264 & 5s\\
Stream2 & 235kbps & 426x240 & video & x264 & 5s\\
Stream3 & 128kbps & 48000Hz & audio & aac & 5s\\
\hline
\end{tabular}
\end{table}

After the encoding, \textit{FFmpeg} was also used to manipulate the selected video and audio segments to create the test sequences. As a first step the needed segments for each quality level of the individual test sequence had to be concatenated as follows:
\begin{lstlisting}[basicstyle=\small, breaklines=true, language=Bash]
$ cat init.m4s seg-1.m4s seg-2.m4s >> firstVideoPart.m4s
\end{lstlisting}
The cat command is used to read the data of the following files and to write them in descending order to the defined output file. In case the output file does not exist it will be created. The number of segments that have to be concatenated depends on the individual segment length as well as the length and design of the test sequences. In this study, test sequences were set to 30 seconds and the segment length during encoding was set to 5 seconds. Therefore, all test sequences that were designed to display the stalling event, or switch, in the middle of the test sequence required concatenating four segments. The initialization segment is followed by three segments each representing 5 seconds of content, totaling a length of 15 seconds per concatenation. The test sequences for the multiple quality switch, which represented a different temporal event pattern, only required three segments per concatenated file. Again the initialization segment is followed by two segments resulting in an output representing 10 seconds of content.
After concatenating, the output file had to be transformed to \textit{.mp4} file format with the following \textit{FFmpeg} command:
\begin{lstlisting}[basicstyle=\small, breaklines=true, language=Bash]
$ ffmpeg -i firstVideoPart.m4s -c copy firstVideoPart.mp4
\end{lstlisting}

First, the path to the input file is specified and then its data is copied in the described output format, in this case, \textit{.mp4}. The \textit{-c copy} option will prevent \textit{FFmpeg} to re-encode the input data \cite{ffmpegdocumentation}. \\
The process of concatenation and transformation to \textit{.mp4} had to be repeated for every individual part that was necessary to construct the test sequences. The high amplitude quality switch as well as both stalling events were constructed from two individual parts each representing 15 seconds of content. The multiple quality switch, as it displays three different quality levels, was constructed from three individual parts each representing 10 seconds of content. In addition, audio segments had to be concatenated and transformed in the same manner. Once this process had been concluded, \textit{FFmpeg} was used to merge the individual video and audio parts:
\begin{lstlisting}[basicstyle=\small, breaklines=true, language=Bash]
$ ffmpeg -y -i firstVideoPart.mp4 -i firstAudioPart.mp4 -c:v copy -c:a copy firstPart.mp4
\end{lstlisting}
Initially, the y-flag command is used to give global permission to overwrite the output. Then two inputs are defined and mapped, the first representing the video and the second the audio source. As before the copy command is used to ensure that inputs are not re-encoded. Lastly, the output filename and format are defined \cite{ffmpegdocumentation}. This process also had to be repeated until all individual audio and video parts, which were needed for the designed test sequences, were combined.
For the imitation of a stalling event, an additional buffering sequence was needed. To create a realistic scenario \textit{FFmpeg} was used to cut the last frame of the GOP previous to the stall, which was then inserted as the background for the buffering wheel video. This was achieved by the following two commands:

\begin{lstlisting}[basicstyle=\small, breaklines=true, language=Bash]
$ ffmpeg -sseof -3 -i firstStallPart.mp4 -update 1 -q:v 1 lastFrame.png 

$ ffmpeg -y -framerate 24 -loop 1 -i lastFrame.png -i loadingWheel.mp4 -filter_complex 
'[1]format=argb, colorchannelmixer=aa=0.5[ol];[0][ol]overlay' -t 1 finalBufferEffect.mp4
\end{lstlisting}
The first command creates the screen capture of the last frame. The sseof option tells \textit{FFmpeg} to seek and output all frames specified by the position, in this case, three seconds, relative to the end of the input file. To ensure that only the last frame is saved the update option is changed from default 0 to 1. This setting will continuously overwrite the data in the specified output file, ultimately leaving only the last frame. The encoding quality is set to 1, which represents the highest quality level. The captured frame is set to be in \textit{.png} file format \cite{ffmpegdocumentation}.
The second command inserts the previously captured frame into the buffering wheel video. The framerate of the output is set to 24 frames per second and the image file is specified to loop throughout the length of the output file. Additionally, options to create a complex filtergraph are defined. First, the color model of the second input referred to as indexed 1, is set to Alpha Red Green Blue (ARGB). Then its value for the alpha channel is set to 0.5, which makes the input source semi-transparent \cite{listofffmpeg}. Following, the first input, marked as indexed 0, is set as the main input on which the second input will be overlaid. Then output length is set to one second and the output name and format are defined \cite{ffmpegdocumentation}.
Once the buffering sequence for all content types was created, all previously prepared parts were concatenated with \textit{FFmpeg} to create the final test sequences. Therefore the following command was used:

\begin{lstlisting}[basicstyle=\small, breaklines=true, language=Bash]
$ ffmpeg -y -f concat -safe 0 -i list.txt -c copy finalTestSequence.mp4
\end{lstlisting}
First, global overwrite permission was defined. The f-flag option enforces the file format of the concatenated sequences on the output. The safe option was set to 0 to disable safe mode, which allows the input of all files. The input specifies the path to a \textit{.txt} file which contains the relative paths to the required files for the test sequences in concatenation order. As an example the structure of a \textit{.txt} file used for the two-second stall test sequence is given:
\begin{lstlisting}[basicstyle=\small, breaklines=true, language=Bash]
file 'stalling_event/firstPart.mp4'
file 'stalling_event/finalBufferEffect.mp4'
file 'stalling_event/finalBufferEffect.mp4'
file 'stalling_event/secondPart.mp4'
\end{lstlisting}
Lastly, the input data is copied without further encoding to the specified output \textit{.mp4} format. This concatenation process had to be repeated for each of the defined test sequences of this study \cite{ffmpegdocumentation}.

\subsection{Effect Design}
\label{effect}
\begin{figure*}
    \begin{subfigure}[b]{0.23\linewidth}
        \centering
        \includegraphics[width=\textwidth]{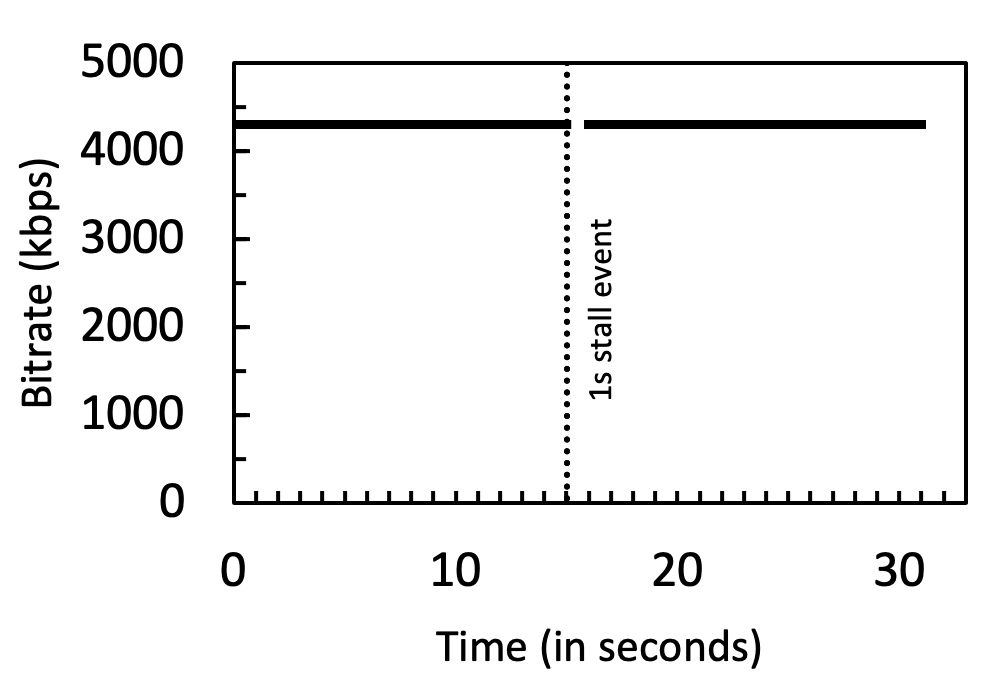}
        \caption{\footnotesize 1s stall event at 15s}
    \end{subfigure}
    \hfill
    \begin{subfigure}[b]{0.23\linewidth}
        \centering
        \includegraphics[width=\textwidth]{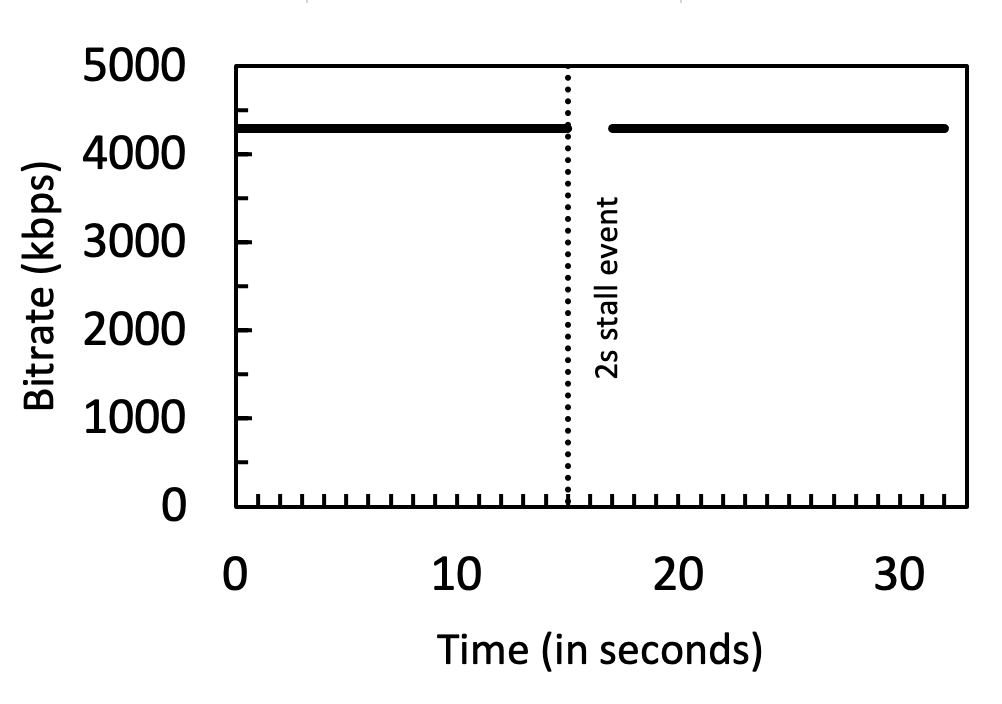}
        \caption{\footnotesize 2s stall event at 15s}
    \end{subfigure}
    \hfill
    \begin{subfigure}[b]{0.23\linewidth}
        \centering
        \includegraphics[width=\textwidth]{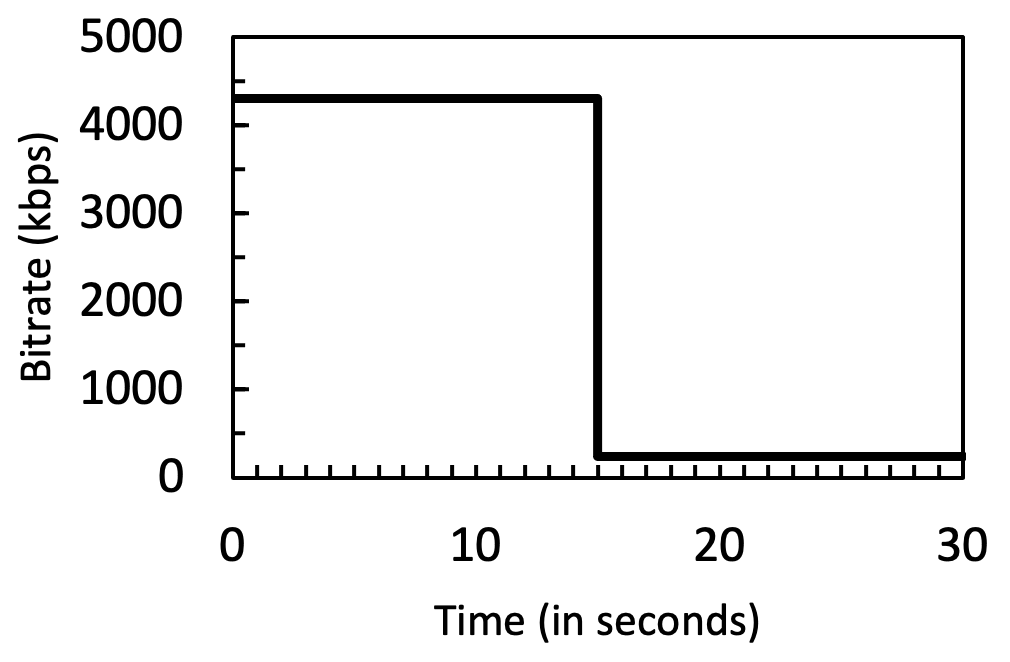}
        \caption{\footnotesize High amplitude quality switch}
    \end{subfigure}
    \hfill
    \begin{subfigure}[b]{0.23\linewidth}
        \centering
        \includegraphics[width=\textwidth]{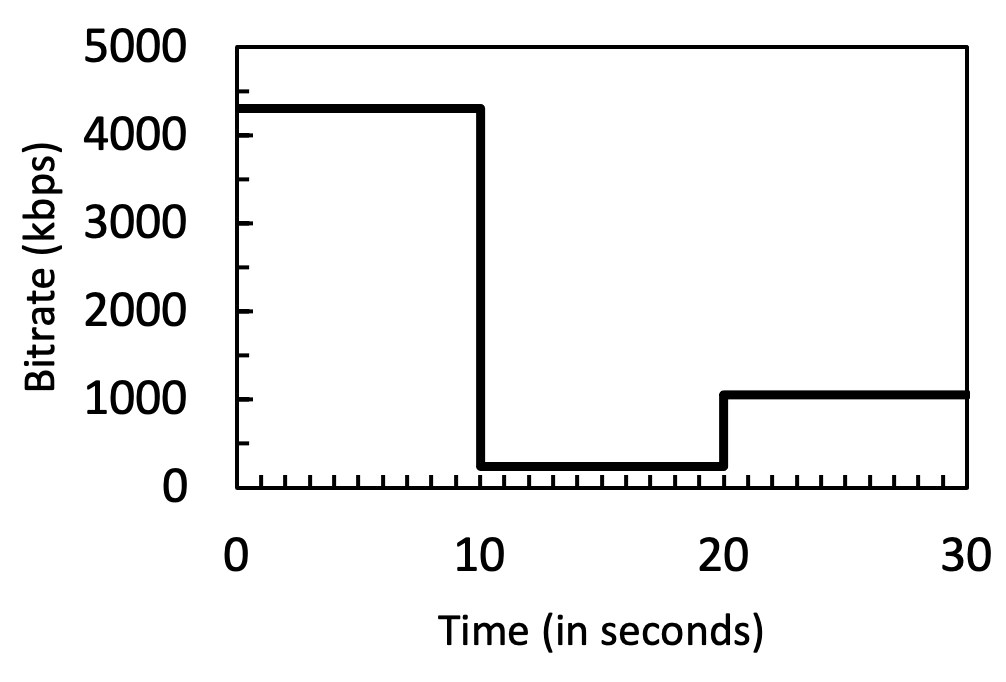}
        \caption{\footnotesize Multiple quality switch}
    \end{subfigure}
    \caption{Representation of the One and Two Second Stall Events and High Amplitude Quality Switch (1080p-240p) and Multiple Quality switches (1080p-240p-480p)}
    \label{fig:os}
\end{figure*}

To answer the research question and to measure the different effects of stalling events, multiple quality switches and high amplitude quality switches on QoE the following effects are proposed: 1-second stalling event Figure~\ref{fig:os}(a), 2-second stalling event Figure~\ref{fig:os}(b), a high amplitude quality switch from 1080p to 240p Figure~\ref{fig:os}(c), and a multiple quality switch from 1080p to 240p to 720p Figure~\ref{fig:os}(d). The decision for introducing the 1-second stall was based on a study by De Pessemier~\etal, which quantified the effects of stalling events \cite{pessemier}. Within this study, the median of a single stall event fluctuated around one second. The 2-second stall event was chosen to analyze the effect difference between a 1-second stall compared to a 2-second stall. The quality levels for the high amplitude as well as the multiple quality switch were based on a study that investigated, in particular, the impact of different switching patterns on QoE \cite{switchmain}. The study indicated that both events would resolve in worse QoE than other events. To avoid additional influences of effects such as the primacy or recency effect, which assume that the first or last seconds of a video has a higher impact on the viewers' QoE, the stall event was placed in the middle of each test sequence \cite{recency}.

\subsection{Graphical User Interface}
\label{interface}
In order to display the study to the participants, a website-based Graphical User Interface (GUI) was developed. The website was based on HTML, CSS, and JavaScript. IndexedDB, a low-level API that can pre-cache data on the client-side, was used to prevent unwanted impairments \cite{indexeddbmdn}. The caching process of the test sequences would immediately commence when the participant opened the website and continued in the background without any disturbance. This was implemented to reduce potential waiting times for the users later in the study. In the backend, a node.js express server was used to request the prepared test sequences from the hosting server and to handle the connection to the database. The test sequences were stored using Amazon S3 and, from there, requested to the website. For deployment, Amazon's EC2 instances were chosen. In accordance with the best practice framework, consideration was given to the size of the instance. Ultimately, a small instance type was chosen, which was expected to be sufficient for the extent of this study. Upon opening the website, participants would see an introductory text that explained the study procedure and an input field that asked for an MTurk ID. After submission, an example video sequence would be displayed to familiarize the participants with the player design. The video player, implemented using the HTML video tag, was minimized to the viewing field and all additional customization options such as fullscreen, start/stop, and audio options were deactivated. Furthermore, the width of the video player was fixed to 1280px. This measure was taken to ensure that every participant would have the same experience from a technical perspective. A customized overlay was cast on each test sequence to allow participants to pause if necessary. Once the playback of a test sequence was started, the video player would position itself in the middle of the screen and the background would turn grey as suggested in the ITU-T standard P.910 \cite{itutstudy}. After the test sequence finished, the background would turn white, and only then would the questionnaire appear. This functionality was implemented to ensure that participants could not simply click through the study without watching the test sequences. As suggested in \cite{kittur2008crowdsourcing} this measure would also ensure that the proper completion of the study would require the same time as a cheating attempt. Each questionnaire contained a content-related question, which was treated as the gold standard data and the 5-point quality scale scores.

\begin{figure*}
    \centering
    \includegraphics[width=\linewidth]{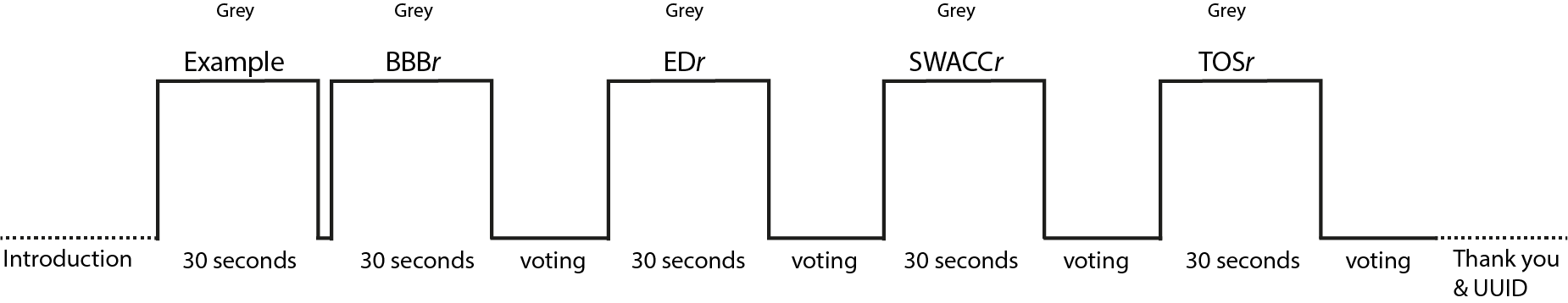}
    \caption{Graphical Representation of the ACR Stimulus Presentation in the Subjective Study - BBB\textit{r}, Big Buck Bunny test sequence at random effect \textit{r} - ED\textit{r}, Elephants Dream test sequence at random effect \textit{r} - SWACC\textit{r}, The Swiss Account test sequence at random effect \textit{r} - TOS\textit{r}, Tears of Steel test sequence at random effect \textit{r}}
    \label{fig:stimulus}
\end{figure*}

As the main aim of this study was set on comparing the resulting quality of the chosen effect types, the ACR methodology was chosen. The test sequences were shown in the following order: Big Buck Bunny, Elephants Dream, The Swiss Account, and Tears of Steel \mbox{(Figure~\ref{fig:stimulus})}. While the content order remained the same between participants, the effects were randomly chosen by an algorithm implemented with JavaScript. The algorithm varied combinations of effect and content types but ensured that each participant would watch each effect once. Upon completion, the results would be sent to the express server and stored in the MongoDB database. The following data was stored: MTurk ID, timestamp of submission, answers to the questionnaire, and an automatically generated UUID for each user. The UUID would also be sent back to the website and displayed to the participant. Mobile, as well as tablet devices, were excluded from the study. If a participant tried to access the website via these devices, an error message would appear suggesting to switch to a desktop or laptop device. Similarly, any browser other than Google Chrome was excluded from the study as trials showed that they did not support the manual bitrate switching within the manipulated sequences. A similar error message would be displayed in this case.

\subsection{Amazon Mechanical Turk}
\label{mturk}
Amazon Mechanical Turk (MTurk) is a crowdsourcing marketplace which enables companies and individuals to outsource tasks virtually \cite{mturk}. The platform allows for unlimited campaign creation and customized settings. Each campaign must have a short description text, a set time frame, and allocated compensation. Other elements that can be customized are among other geographical locations and minimum user rating. A trial campaign with 5 test subjects was created to test functionality and user reliability. The timeframe was set to 15 minutes and compensation \mbox{to 0,50 \$}. To receive compensation, users had to input their MTurk ID in the study's interface as well as paste the UUID which they received after completion back on the MTurk platform. This measure was implemented to ensure that users did complete the study. It also allowed cross-checking between the database and the MTurk platform to filter out double participation.
The initial test confirmed the functionality of the campaign. Despite the given functionality of the interface and the MTurk setup, none of the five datasets provided accurate answers to the gold standard data. For further testing purposes, a second campaign with 10 participants was started. In addition to the prior settings, this campaign was only accessible to users with an acceptance rating of 95\% and higher and limited to 10 minutes. Upon initial analysis, the second campaign appeared successful as the gold standard data was answered correctly. Therefore, it was assumed that participants in the second campaign watched the content properly. Hence a third and last campaign with 60 additional participants and the same settings as prior was initiated.

During the in-depth analysis of the MTurk results, the following issues could be discovered: For one, all effects achieved relatively high grading. Secondly, there were many instances in which the participants rated all effect types equally, indicating no difference between effect types.
Based on these findings, the following assumptions were made: a 5-point scale does not give participants enough variety to judge the effects with a meaningful difference. Some participants were only interested in concluding the survey as fast as possible to receive the incentive. Therefore, they did not take the time to properly watch and evaluate the test sequences, so no differences in effect types are noted.

\subsection{Follow-Up Study}
\begin{figure}[tp]
  \centering
  \includegraphics[width=\columnwidth]{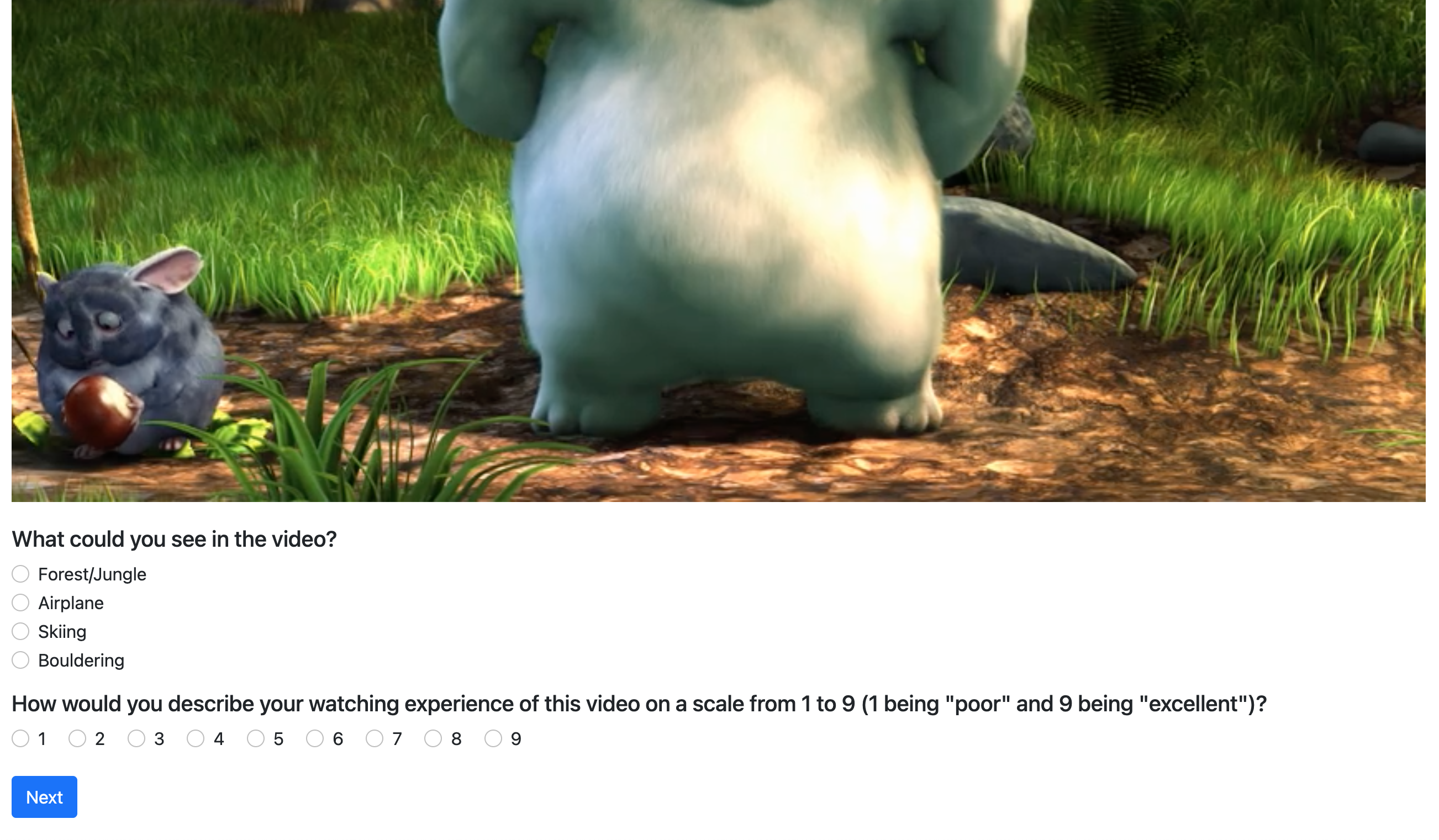}
  \caption{GUI after Test Sequence Playback displaying Questionnaire}
  \label{fig:guiquestionnaire}
\end{figure}

In order to respond to the stated observations, a follow-up study was designed. Three major changes were made to consider the previous findings and assumptions. First of all, the 5-point rating scale was changed to the recommended 9-point scale to give the participants a higher discriminative rating power. Secondly, to respond to the equal ratings between effect types, four additional test sequences were introduced. Therefore, every participant would rate 8 sequences and experience each effect type twice. This change introduced an additional reliability measure as participants who displayed a significant variance within an effect type could be further investigated. 
The additional sequences were taken from the same four movies to prevent further potential content effects. The last change concerned the participants. Participants in the follow-up study received no compensation. Instead, they were taken from a pool of volunteers to diminish the chances of unreliable results due to an incentive. To replace MTurk's time frame setting two additional variables were introduced and saved in the database: timestamp upon watching the first test sequence and timestamp of submission. The introduction of the rating scale on the introductory page and the scale beneath each test sequence was adapted accordingly Figure~\ref{fig:guiquestionnaire}.

\begin{figure*}
    \centering
    \includegraphics[width=\linewidth]{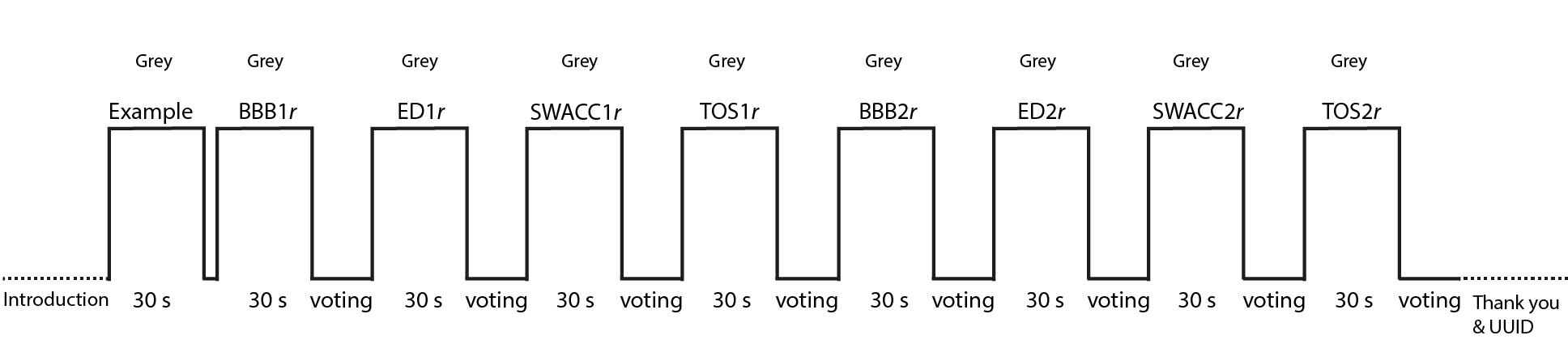}
    \caption{Graphical Representation of the ACR Stimulus Presentation in the Follow-Up Study - BBB1\textit{r} \& BBB2\textit{r}, First and Second Big Buck Bunny test sequence at random effect \textit{r} - ED1\textit{r} \& ED2\textit{r}, First and Second Elephants Dream test sequence at random effect \textit{r} - SWACC1\textit{r} \& SWACC2\textit{r}, First and Second The Swiss Account test sequence at random effect \textit{r} - TOS1\textit{r} \& TOS2\textit{r}, First and Second Tears of Steel test sequence at random effect \textit{r}}
    \label{fig:stimuluspersonal}
\end{figure*}

\section{Statistical Analysis and Findings}
\label{section:results}
The following section will summarize the statistical results for both studies, once with participants which received an incentive and, secondly, with volunteers. For the statistical analysis, a One-Way ANOVA was conducted to validate significant variances between the different effect groups. For the analysis within effect types, a two-sample t-test was performed. All statistical tests used an alpha level of 0,05. Two POST HOC tests were used to validate the significant difference. For one, the Bonferetti test with a corrected alpha value. Secondly, the Tukey B test. In addition to significant variances between effect types, the results were also checked for differences between content types. For reasons of simplicity, figures and tables in the following chapter will describe the effects with the following abbreviations: 1-second stall as \textit{Os}, 2-second stall as \textit{Ts}, multiple quality switch as \textit{Ms}, and high amplitude quality switch as \textit{Ha}.

\subsection{MTurk Evaluation}
In total, 72 entries were submitted to the database. It can be assumed that the additional two participants completed the study but did not manage to input the UUID back on MTurk. Therefore, they were not counted as participants by MTurk but still submitted their results to the database. For this reason, these two as well as 10 other datasets were excluded from analysis due to wrong gold standard data or double participation. The ratings of the remaining 60 participants were then checked for significance with a one-way ANOVA. The evaluation, as seen in Table~\ref{tab:anova} showed a minor significance between the effect groups. Therefore two POST HOC tests were conducted in which all effects were weighed against each other (Table~\ref{tab:posthoc}). Following the Bonferroni method, the corrected alpha value was calculated, resulting in  $\alpha=0,008333$, which indicates no significance between the effect groups. The same results were achieved with the Tukey B POST HOC test with a Q-value of $Q=3,66$ and a resulting T-value of $T=0,41583$.

\subsection{Follow-up Study Evaluation}
For the follow-up study, 43 data sets were collected. According to the ITU-T P.910 recommendation, sample sizes larger than 40 merely result in minimal differences  \cite{itutstudy}. Therefore, the 43 datasets were assumed sufficient for statistical purposes. None of the participants in the follow-up study had any prior experience with video or image ratings. The dataset analysis showed that all participants answered the gold standard data correctly. However, 6 participants were excluded as their study completion time varied two standard deviations from the mean, indicating that they were not paying close attention to the test. A two-sample t-test was performed for each effect type to check for significant differences between the first rating of an effect contrary to the second rating. No significant difference was found. According to the resulting MOS, effects ranked in the following descending order: 1-second stall, multiple quality switch, 2-second stall, and high amplitude quality switch (Figure~\ref{fig:effectAndContent}). The results of the 9-point scale were mapped to a 5-point MOS based on the following formula, where \textit{NP} represents the rating value on the 9-point scale:
\begin{equation*}
    MOS = \frac{NP}{2}+ 0,5
\end{equation*}

Equally to the previous study, the variance between the effect types was checked with a One-Way ANOVA, which indicated a significant variance (Table~\ref{tab:anova}). The corrected alpha remained the same for the Bonferroni POST HOC method with $\alpha=0,008333$. In the case of the 1-second stall and the high amplitude quality switch, as well as the multiple quality switch and the high amplitude quality switch, $P <$ \textit{corrected}  $\alpha=0,00833$, indicating significance. The Tukey B POST HOC test confirmed these results as $T >$ \textit{Absolute Mean Difference} (Table~\ref{tab:posthoc}). Therefore it can be assumed that there was a significant difference between the high amplitude quality switch and 1-second stall and the high amplitude quality switch and the multiple quality switch events. In both cases, the high amplitude quality switch performed significantly worse.
In addition, the results were checked for significant variance between content types. The one-way ANOVA indicated significance (Table~\ref{tab:anova}). The results of the Tukey B POST \mbox{HOC ($T = 0,764$)} confirmed a significant difference between Big Buck Bunny and the other movie sequences. Throughout all effect types, the Big Buck Bunny clips achieved the highest MOS (Figure~\ref{fig:effectAndContent}).

\begin{table}[tp]
  \caption{Analysis of Variance (One-Way ANOVA, between Groups) of the different Effect and Content Types of the Subjective Results of MTurk and Laboratory Participants}
  \label{tab:anova}
  \begin{threeparttable}
  \begin{tabular}{|c|c|c|c|c|c|}
    \hline
    \textbf{Evaluation} & \textbf{df}~\tnote{1} & \textbf{Mean Square} & \textbf{F}~\tnote{2} & \textit{\textbf{P}}~\tnote{3} & \textbf{F}~\tnote{4}\\
    \hline
    Effects (MTurk) & 3 & 2.070 & 2.673 & 0.048 & 2.642\\ 
    Contents (MTurk) & 3 & 1.348 & 1.720 & 0.163 & 2.642\\
    \hline
    Effects (Lab) & 3 & 4.721  & 5.758 & 0.00077 & 2.635\\ 
    Contents (Lab) & 3 & 22.382  & 6.900 & 0.00016 & 2.635\\ 
    \hline
  \end{tabular}
  \begin{tablenotes}
    \item[1] Degree of freedom, number of total categories minus 1
    \item[2] F-value, the significance of variance between two populations
    \item[3] P-value, significant if $P<0.05$ (chosen $\alpha$ level)
    \item[4] Critical F-value, if $F>Fcrit$ then significance
  \end{tablenotes}
  \end{threeparttable}
\end{table}

\begin{table}[tp]
  \caption{Post hoc (Bonferroni, corrected $\alpha=0,0083$) and Tukey B ($T=0,415$) for MTurk and Laboratory Participants between Effect Types}
  \label{tab:posthoc}
  \begin{threeparttable}
  \begin{tabular}{|c|c|c|c|c|c|c|}
    \hline
    \textbf{Environment} & \textbf{I}~\tnote{1} & \textbf{J}~\tnote{1} & \textbf{P}~\tnote{2} & \textbf{Absolute Mean Difference}~\tnote{3}\\
    \hline
    MTurk & Ha & Ms & 0.649 & 0.083 \\ 
    MTurk & Ha & Os & 0.066 & 0.3 \\ 
    MTurk & Ha & Ts & 0.018 & 0.4 \\ 
    MTurk & Ms & Os & 0.161 & 0.216 \\ 
    MTurk & Ms & Ts & 0.049 & 0.316 \\ 
    MTurk & Os & Ts & 0.458 & 0.1 \\
    \hline
    Laboratory & Ha & Ms & 0.00099 & 0.486 \\ 
    Laboratory & Ha & Os & 0.00012 & 0.574 \\ 
    Laboratory & Ha & Ts & 0.0253 & 0.358 \\ 
    Laboratory & Ms & Os & 0.526 & 0.087 \\ 
    Laboratory & Ms & Ts & 0.399 & 0.128 \\ 
    Laboratory & Os & Ts & 0.159 & 0.216 \\ 
    \hline
  \end{tabular}
  \begin{tablenotes}
    \item[1] I \& J, different effect types
    \item[2] Two-tail t-test
    \item[3] $\mid I - J \mid$
  \end{tablenotes}
  \end{threeparttable}
\end{table}

\section{CONCLUSIONS AND FUTURE WORK}
\label{section:conclusions}
This paper investigated whether stalling events have the highest impact on decreasing the viewer’s QoE compared to multiple or high amplitude quality switches. The results confirm that when compared to multiple and high amplitude quality switches, stalling events do not have the worst impact on QoE. The results suggest ABR algorithms should consider introducing short stall events instead of high amplitude quality switches. However, further studies with different stalling durations, switching patterns, and quality switching amplitudes are advised.

\section*{Acknowledgment}
The financial support of the Austrian Federal Ministry for Digital and Economic Affairs, the National Foundation for Research, Technology and Development, and the Christian Doppler Research Association is gratefully acknowledged. \url{https://athena.itec.aau.at/}.

\bibliographystyle{IEEEtran}
\bibliography{IEEEabrv, ref}

\end{document}